\documentclass[journal]{IEEEtran}
\pdfoutput=1
\usepackage{color}
\usepackage{soul}
\usepackage{afterpage}
\usepackage{url}
\usepackage{ulem}
\usepackage{hyperref}
\usepackage{amsmath}
\usepackage{graphicx}
\usepackage[percent]{overpic}
\newlength{\deffigheight}
\newcommand{\afig}[4]{%
\begin{overpic}[#1]{#2}%
\put(#3){#4}%
\end{overpic}}

\begin{document}

\title{NbTi/Nb/Cu multilayer shield for the superconducting shield (SuShi) septum}

\author{\IEEEauthorblockN{D.~Barna\IEEEauthorrefmark{1}\IEEEauthorrefmark{2}, M.~Nov\'ak\IEEEauthorrefmark{1}, K.~Brunner\IEEEauthorrefmark{1}, C.~Petrone\IEEEauthorrefmark{3}, M.~Atanasov\IEEEauthorrefmark{3}, J.~Feuvrier\IEEEauthorrefmark{3}, M.~Pascal\IEEEauthorrefmark{3}}\\
  \IEEEauthorblockA{\IEEEauthorrefmark{1}MTA Wigner Research Centre for Physics, Budapest}\\
  \IEEEauthorblockA{\IEEEauthorrefmark{2}barna.daniel@wigner.mta.hu}\\
  \IEEEauthorblockA{\IEEEauthorrefmark{3}CERN, Geneva}}

\IEEEoverridecommandlockouts
\IEEEpubid{\makebox[\columnwidth]{978-1-5386-5541-2/18/\$31.00~\copyright2018 IEEE \hfill} \hspace{\columnsep}\makebox[\columnwidth]{ }}

\maketitle

\IEEEpubidadjcol

\begin{abstract}
A passive  superconducting shield  was proposed  earlier to  realize a
high-field (3-4  T) septum  magnet for  the Future  Circular Collider.
This paper  presents the  experimental results  of a  potential shield
material,  a NbTi/Nb/Cu  multilayer sheet.   A cylindrical  shield was
constructed from two halves, each consisting  of 4 layers with a total
thickness of 3.2~mm, and inserted into  the bore of a spare LHC dipole
corrector magnet (MCBY).  At 4.2~K, up  to about 3.1~T at the shield's
surface  only a  leakage  field  of 12.5~mT  was  measured inside  the
shield.  This can  be attributed to the mis-alignment of  the two half
cylinders, as confirmed by finite  element simulations.  With a better
configuration we estimate  the shield's attenuation to  be better than
$\mathbf{4\times 10^{-5}}$,  acceptable for the  intended application.
Above 3.1~T the  field penetrated smoothly.  Below that  limit no flux
jumps were observed  even at the highest achievable ramp  rate of more
than 50~mT/s at  the shield's surface.  A 'degaussing'  cycle was used
to eliminate the effects of the field trapped in the thick wall of the
shield, which could otherwise distort the homogeneous field pattern at
the extracted beam's position.  At  1.9~K the shield's performance was
superior to that at 4.2~K, but it suffered from flux jumps.
\end{abstract}


\section{Introduction}

The  Future Circular  Collider (FCC)  study  was launched  in 2014  to
identify the  key challenges of the  next-generation particle collider
of  the post-LHC  era,  propose technical  solutions  and establish  a
baseline design.   In its  early phase the  parameters are  subject to
frequent changes.  The  current values of the  relevant parameters are
shown in  Table~\ref{tab:fcc-parameters}.  One of the  key problems of
the proton-proton ring  is the high beam rigidity and  the very strong
magnetic fields required to manipulate  this beam. A new generation of
superconducting  dipole magnets  using  Nb$_3$Sn  conductors is  being
developed to  produce the  16~T field  necessary to  keep the  beam on
orbit.   The beam  extraction  system uses  so-called septum  magnets,
which create zero field at the position of the circulating beam, and a
high field  region in  close proximity for  the extracted  beam kicked
off-orbit by upstream kicker magnets.  The unprecedented beam rigidity
(a factor of 6.6 higher than  in today's highest-energy ring, the LHC)
puts serious requirements on these  magnets as well.  A magnetic field
of at least  3~T is desired in  order to keep the total  length of the
septa  within  limits,  and   the  apparent  septum  thickness  (total
thickness  of all  materials, including  beam pipes  and beam  screens
between the two  regions) needs to be minimized in  order to relax the
requirements on  the kicker  magnets' strength.   The target  value is
25~mm, which  corresponds to  a thickness  of 17-18  mm of  the shield
itself, without  beam pipes and  beam screens.   These lead to  a very
sharp transition  between the high-field  and no-field regions  of the
septa.  These requirements are even more important for the high-energy
LHC (HE-LHC) option  (an alternative to the FCC), which  would use FCC
technology in the LHC tunnel, where space is very limited.

\begin{table}
  \begin{center}
    \caption{Relevant parameters of the Future Circular Collider}
    \label{tab:fcc-parameters}
    \begin{tabular}{|llrl|}
      \hline
      Parameter & Symbol & Value & Unit \\\hline\hline
      Circumference & & 80-100 & km \\
      Collision energy & & 50+50 & TeV \\
      Injection energy & & 1.3/3.3 & TeV \\
      Septum field homogeneity & & $\pm$1.5 & \% \\
      Septum integrated field & $\int B\,\mathrm{d}l$ & 190 & Tm\\
      Deflection by the septa & $\alpha_s$ & 1.14 & mrad \\
      Deflection by the kickers & $\alpha_k$ & 0.13 & mrad \\
      Maximum apparent septum thickness & & 25 & mm \\
      \hline
    \end{tabular}
  \end{center}
\end{table}

In  a  recent   proposal  \cite{Barna2017HighFieldSeptum}  this  field
configuration   would   be   realized   by  the   combination   of   a
superconducting magnet and a  passive superconducting shield, referred
to as a  superconducting shield (SuShi) septum in  the following.  The
geometry of  the shield and  the magnet  winding need to  be optimized
simultaneously  to give  the  required field  homogeneity outside  the
shield.   While  a  complete   demonstrator  prototype  creating  this
homogeneous  field pattern  would  be a  major  project including  the
design and construction of a special superconducting magnet, different
superconducting  shield  materials can  be  easily  tested in  simpler
setups and existing magnets. These  tests can study the performance of
the shield materials  in general, with special focus  on the following
points:  (i) Maximum  shielded  field with  a  given thickness.   This
defines the apparent septum thickness of the septum magnet for a given
magnetic field. (ii)  Stability against flux jumps, which  lead to the
sudden collapse of  the shielding currents and the  penetration of the
magnetic field  to the  circulating beam.   Besides an  immediate beam
abort,  the  shield  would  need  to  be  heated  above  its  critical
temperature and cooled back in zero field (``thermal reset'') in order
to eliminate the trapped field.  This  is a very long process, leading
to  unacceptably long  deadtimes.  The  shield itself  must be  stable
against  spontaneous flux  jumps, and  external perturbations  such as
energy  depositions  due  to  beam  loss  must  be  minimized  in  the
accelerator at this position. (iii) The septum magnet must be ready to
a beam abort at any time, i.e.  its field level must follow the actual
beam momentum in  the ring, from injection to top  energy.  The shield
must  therefore  safely  support   repeated  magnetic  cycles  between
injection and  top energy  field levels, without  flux jumps.   (iv) A
detection mechanism is needed to  detect a developing flux jump safely
before  the  field  level  starts  to rise  at  the  position  of  the
circulating beam,  so that  the beam  can be  aborted.  With  a 100~km
circumference  of the  ring the  full revolution  time is  333~$\mu$s.
This  is  the minimum  time  requirement  for  an advance  trigger  to
synchronize the extraction  with the next abort  gap.  Including other
delays, a safe time interval is a few milliseconds at least.  (v) Even
if the field does not penetrate to  the interior of the shield at all,
a trapped  magnetic field will remain  in its thick wall  after a high
field   exposure,  which   will   distort   field  homogeneity,   most
significantly at  low external field  levels, i.e.  at  injection into
the ring.  Elimination of this trapped field by a thermal reset is not
possible due  to deadtime  reasons, as argued  above.  The  tests must
demonstrate other possibilities.

The results presented in this paper are an extension of the
work   carried  out   by  \cite{Itoh1993MagneticShielding}   with  a
different shield  configuration that is suitable  for constructing a
septum magnet, and addressing further issues not studied in that paper.

\section{Experimental setup}

\begin{figure}
  \centerline{\includegraphics[width=\linewidth]{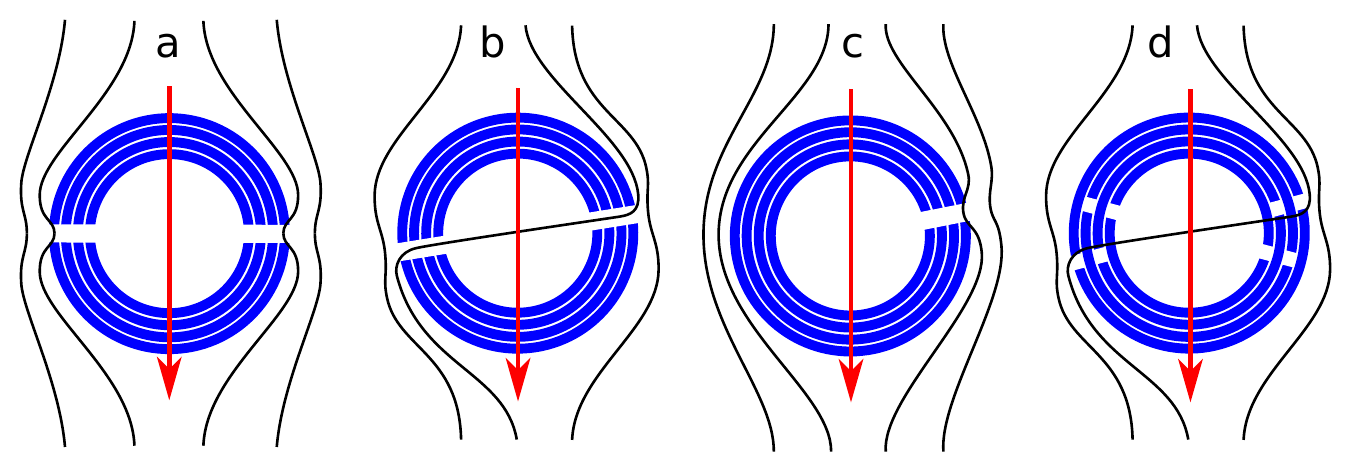}}
  \caption{Different cylindrical shield configurations  made from a
    sheet material, and alignment errors. The vertical red
      arrows indicate the direction of the external dipole field of
      the MCBY magnet in the absence of the shield. Black lines
      indicate induction lines schematically in the presence of the shield. (a) Ideal
      alignment of two half-cylinders with respect to an external
      dipole field. (b) Mis-aligned arrangement of two
      half-cylinders. (c) Mis-aligned arrangement of concentric
      C-shape elements. (d) Zig-zag arrangement of half cylinders. }
  \label{fig:cut-and-exp-layout}
\end{figure}

A cylinder  of length $L$=450~mm, inner/outer  diameter 41/47.4~mm was
constructed from two half cylinders, each  consisting of 4 layers of a
0.8~mm         thick        NbTi/Nb/Cu         multilayer        sheet
\cite{Itoh1993MagneticShielding,Itoh2002NipponTechnicalReport}.   This
material was used earlier for the construction of the inflector magnet
for            the           BNL            $g$-2           experiment
\cite{Yamamoto2002TheSuperconductingInflector},   and   to  create   a
magnetic field concentrator  \cite{Zhang2011ANewStructure}.  The sheet
is the discontinued proprietary product  of Nippon Steel Ltd.  Similar
sheets are currently not available from other vendors. The aim of this
experiment was to confirm  the excellent shielding properties reported
in  \cite{Itoh1993MagneticShielding}, and  test further  aspects which
are important for its application  in a SuShi septum magnet.  Material
R\&D is beyond the scope of  the SuShi septum project, and therefore a
semi-finished  sheet was  purchased from  the remaining  stock of  the
company, and post-processed by the  developer engineer of the sheet in
a   private  company   in   Japan.   Public   information  about   the
manufacturing         process          is         described         in
\cite{Itoh1993MagneticShielding,Itoh2002NipponTechnicalReport}     and
summarized below.  The  sheet was manufactured by packing  NbTi and Cu
sheets alternately into a copper  box, interleaved with thin Nb sheets
at each interface.  The box was closed by electron  beam welding under
vacuum, and then hot rolled, cold  rolled and heat treated.  The final
thickness of the  30 NbTi layers is around 9.5~$\mu$m.   The Cu layers
have the same thickness, except the two outermost ones being 95~$\mu$m
thick. The thickness  of the Nb layers is 0.95~$\mu$m,  and their role
was to prevent the diffusion of  Ti into Cu during the heat treatment.
The NbTi sheets were manufactured by hot forging, hot rolling and cold
rolling of a commercially available Nb-46.5wt\%Ti ingot.  Commercially
available four-9 OFHC  copper (estimated RRR=100) was used  for the Cu
sheets.  Parameters of the heat  treatment have an important effect on
the  critical  current  density  of   the  material,  as  reported  in
\cite{Itoh2002NipponTechnicalReport,Itoh1995CriticalCurrentDensity}.
The  filling factor  of  the composite  by NbTi  is  about 36\%.   The
resulting  multi-layer structure  is  a 2D  analogue  of the  standard
superconductor  cables, where  superconducting filaments  are embedded
typically in a copper matrix.  The NbTi layers are responsible for the
high current  densities and thereby  the shielding performance  of the
material, and Cu is used for stabilization.

\begin{figure*}
   \centerline{\includegraphics[width=0.444\linewidth]{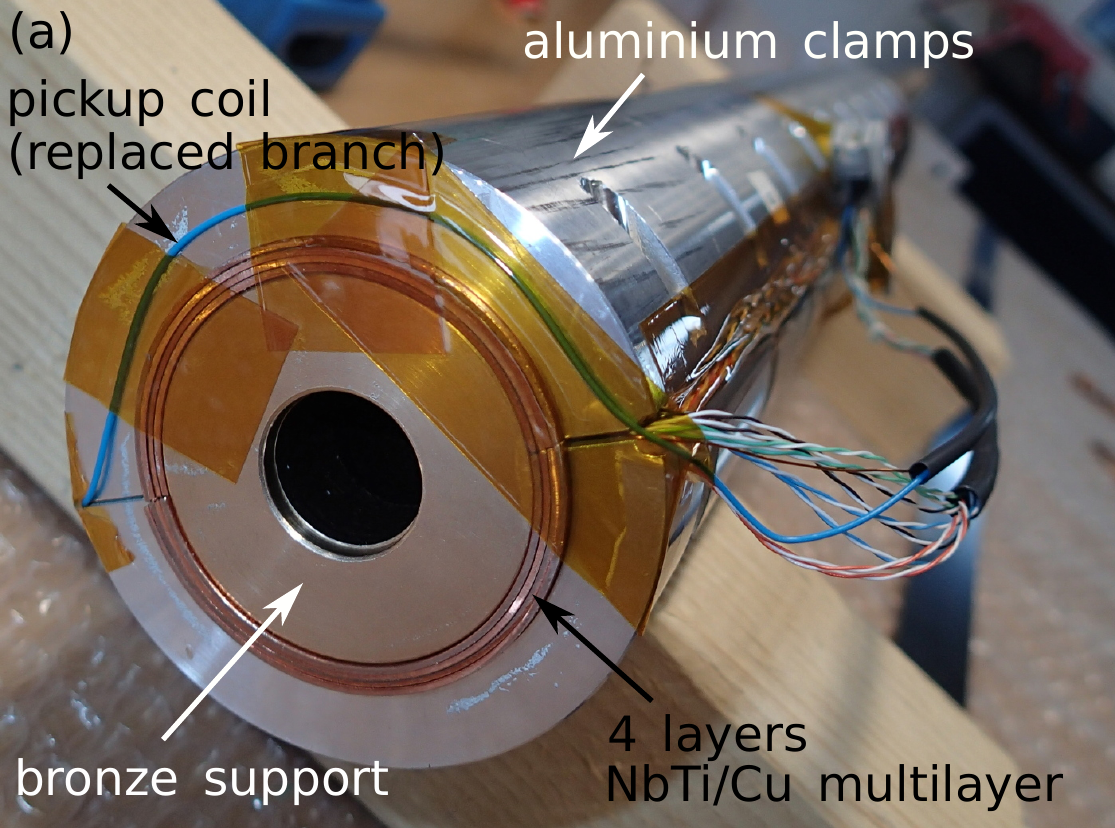}\hspace*{\fill}%
     \includegraphics[width=0.525\linewidth]{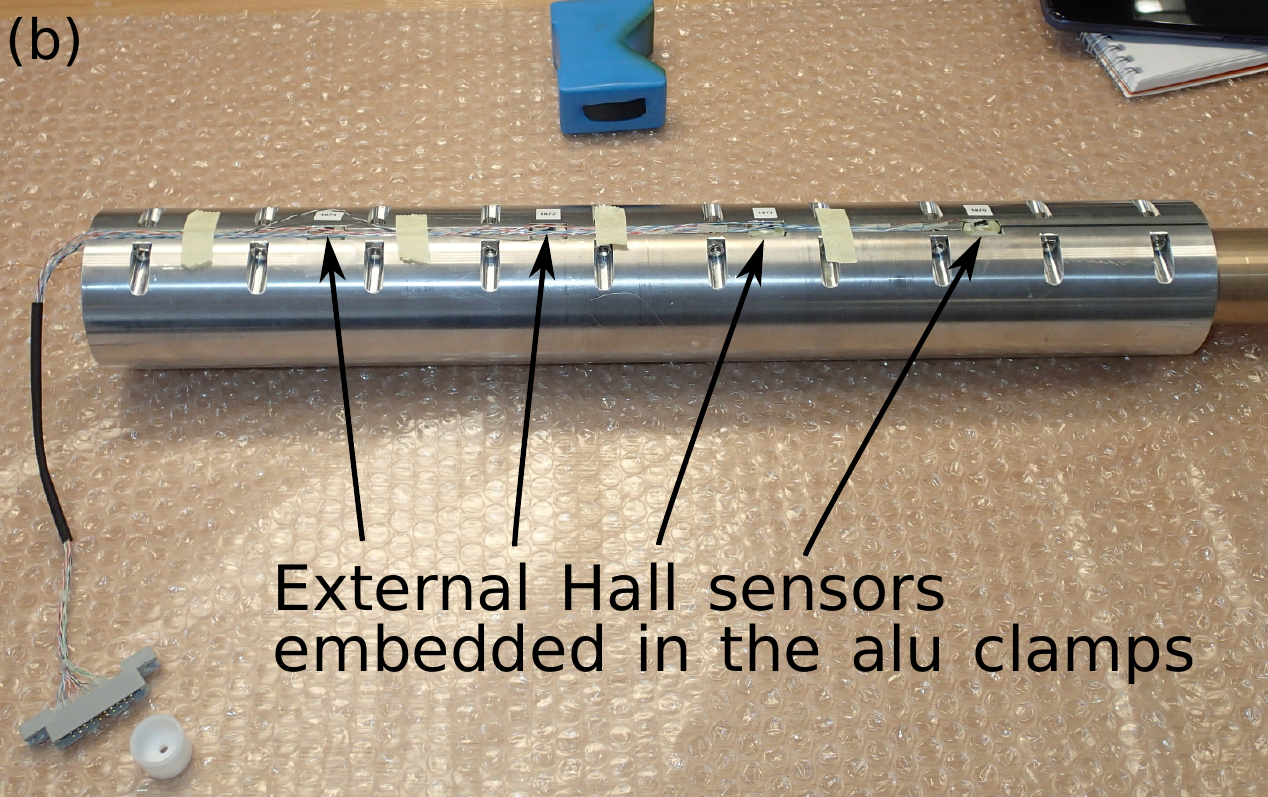}}
\caption{The superconducting shield  assembly. (a) End view
    showing the construction of the assembly and the pickup coil. (b) Positions of the external Hall sensors.}
\label{fig:superconducting-shield-construction}
\end{figure*}

Figures~\ref{fig:cut-and-exp-layout}(a-d)  show different  cylindrical
shield configurations for a transverse  dipole field which can be made
from a sheet material.  With two half cylinders aligned perfectly with
respect to the external field (a)  the shielding currents do not cross
the plane of the cut, and the leaking magnetic field inside the shield
is  parallel  to  the  external  field.  If  the  shield  is  slightly
misaligned (b), induction lines can pass  through the two cuts and the
major  component  of the  leak  field  will  be perpendicular  to  the
external  field.    This  effect  is  eliminated   and  the  shielding
efficiency is  made less sensitive  to misalignments if the  shield is
made    from    concentric    C-shape    elements    as    shown    in
Fig.~\ref{fig:cut-and-exp-layout}(c).   An alternating  arrangement of
the cuts  on the  left and  right sides of  configuration (c)  is even
better.  Even though the configuration  (c) was planned initially, the
sheets were  accidentally cut to  half without excess  material, which
finally  only  allowed  the  realization  of  the  two  half-cylinders
configuration  (b), without  the  possibility to  machine the  meeting
sides of the half cylinders to a flat surface.  A  further
  possible  configuration   with  half-cylinders  is   illustrated  in
  Fig.~\ref{fig:cut-and-exp-layout}(d).    Whether  this   arrangement
  improves the shielding efficiency  with respect to configuration (b)
  is a function of the degree of misalignment of the latter.  Although
  configuration (d) seems  to be symmetric in  average, the subsequent
  layers from inside to outside  are exposed to an increasing magnetic
  field, and therefore  the effects of their  rotations have different
  weights, leading to the schematic field pattern shown in the figure,
  confirmed by  finite element simulations. Although  for large enough
  misalignments  of  configuration  (b) the  configuration  (d)  could
  perform better, our strategy was  to assemble configuration (b) with
  the best possible alignment.   Mounting configuration (d) would also
  have been  difficult due to  the spring-back effect of  the shells.
In  the  final  setup  there  remained  small  gaps  between  the  two
half-cylinders.   In  addition,  the  different layers  could  not  be
perfectly  aligned during  the assembly.   The cuts  in the  different
layers  had slightly  different  orientations, also  varying with  the
axial position.  The  tested configuration is therefore  that shown in
Fig.~\ref{fig:cut-and-exp-layout}(b).

\begin{figure}
  \centerline{\includegraphics[width=\linewidth]{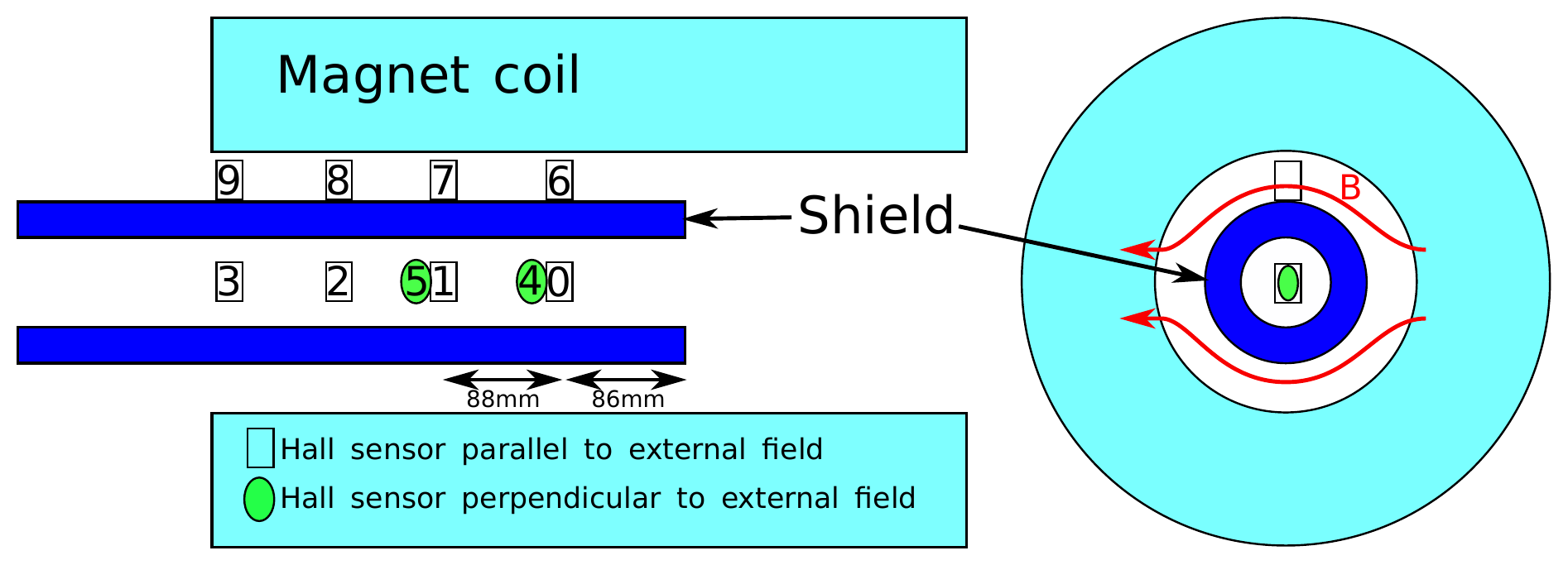}}
  \caption{Schematic layout of the experimental setup and the
    positions and numbering of the Hall sensor slots. Not to scale!
    The red arrows indicate the induction lines in the presence of the
  shield.}
  \label{fig:schematic-experimental-layout}
\end{figure}

\begin{figure*}
  \newlength{\figheight}
  \setlength{\figheight}{5.0cm}
  \centerline{%
    \includegraphics[height=\figheight]{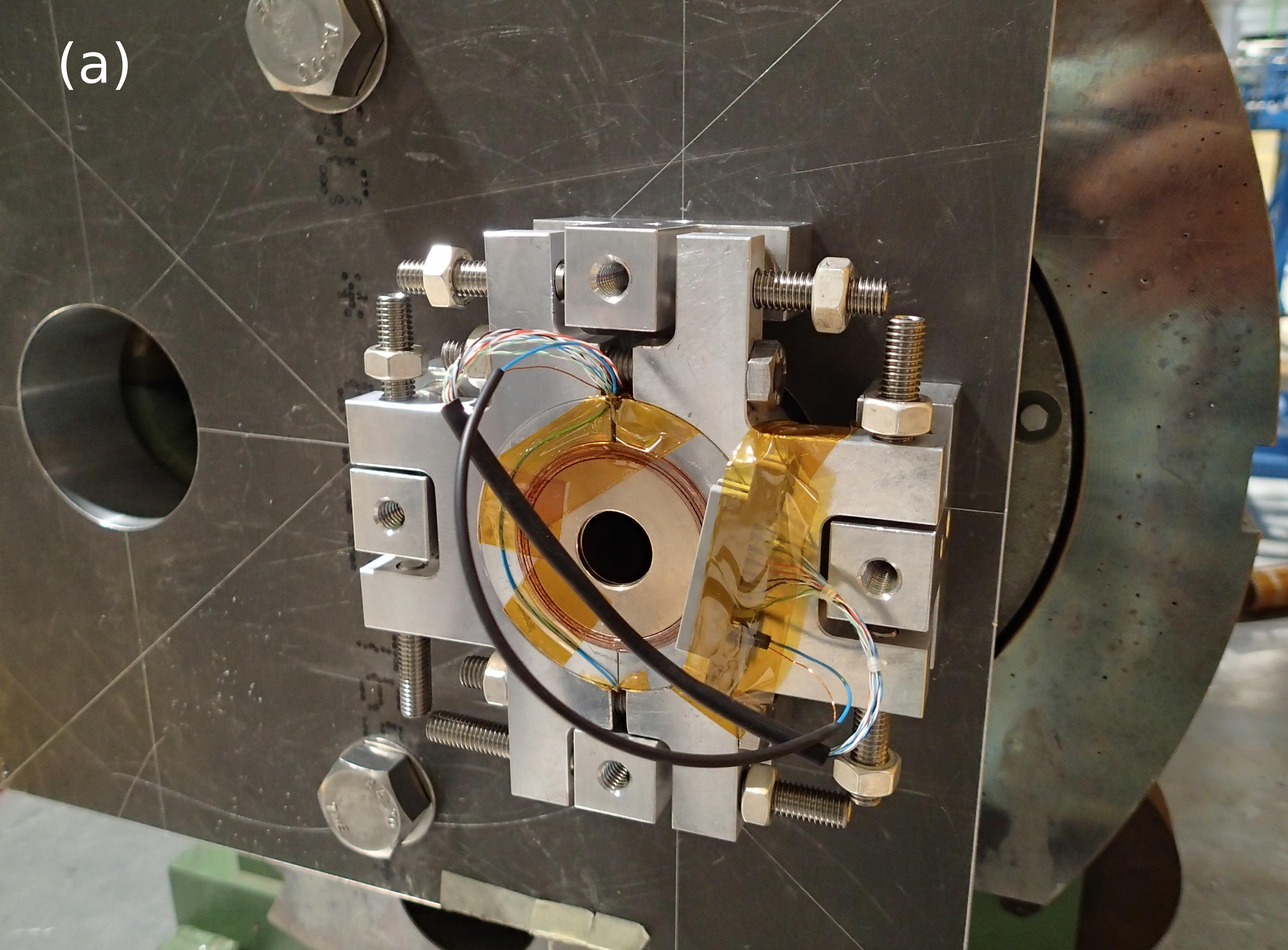}%
    \hspace*{\fill}%
    \includegraphics[height=\figheight]{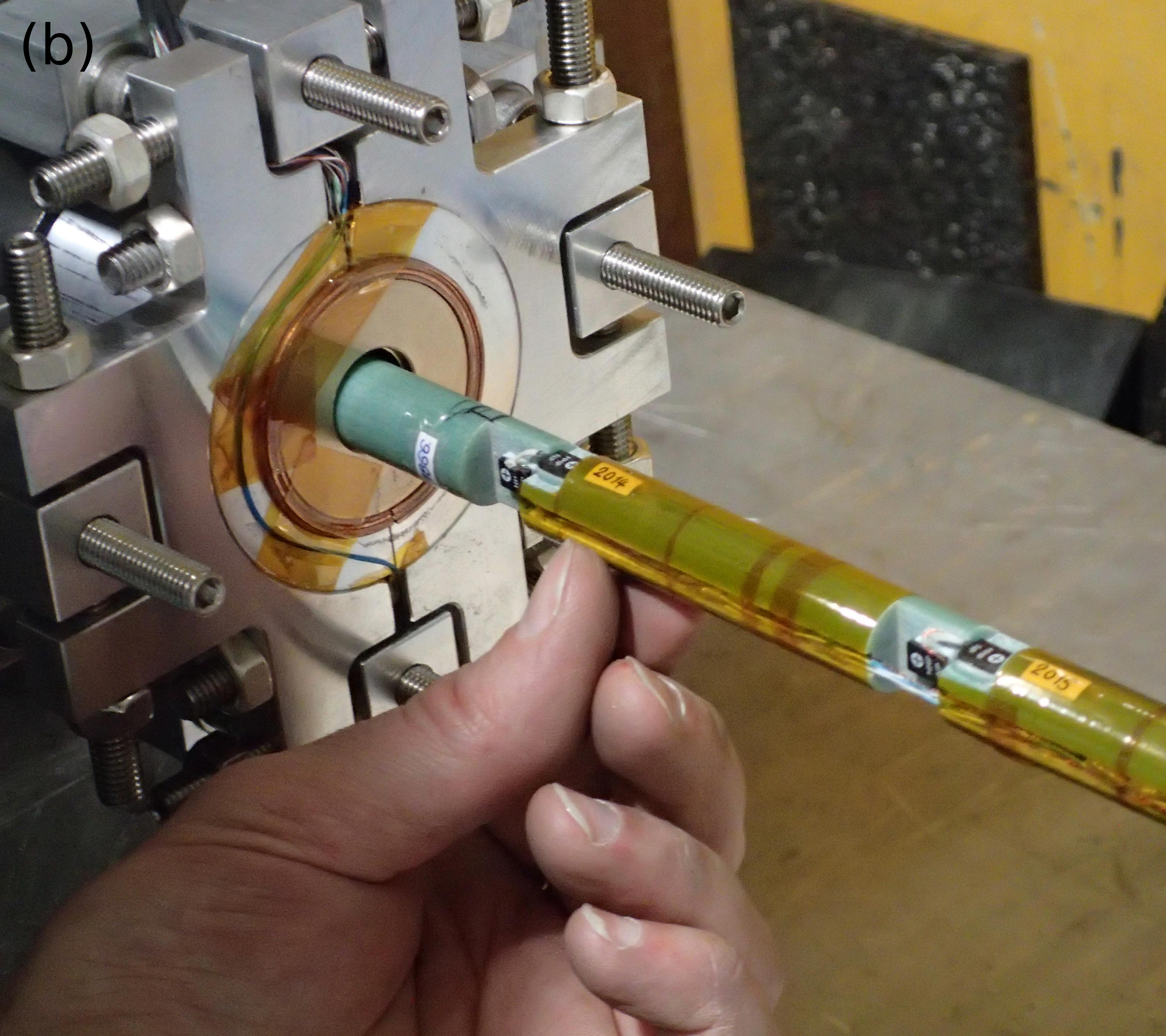}%
    \hspace*{\fill}%
    \includegraphics[height=\figheight]{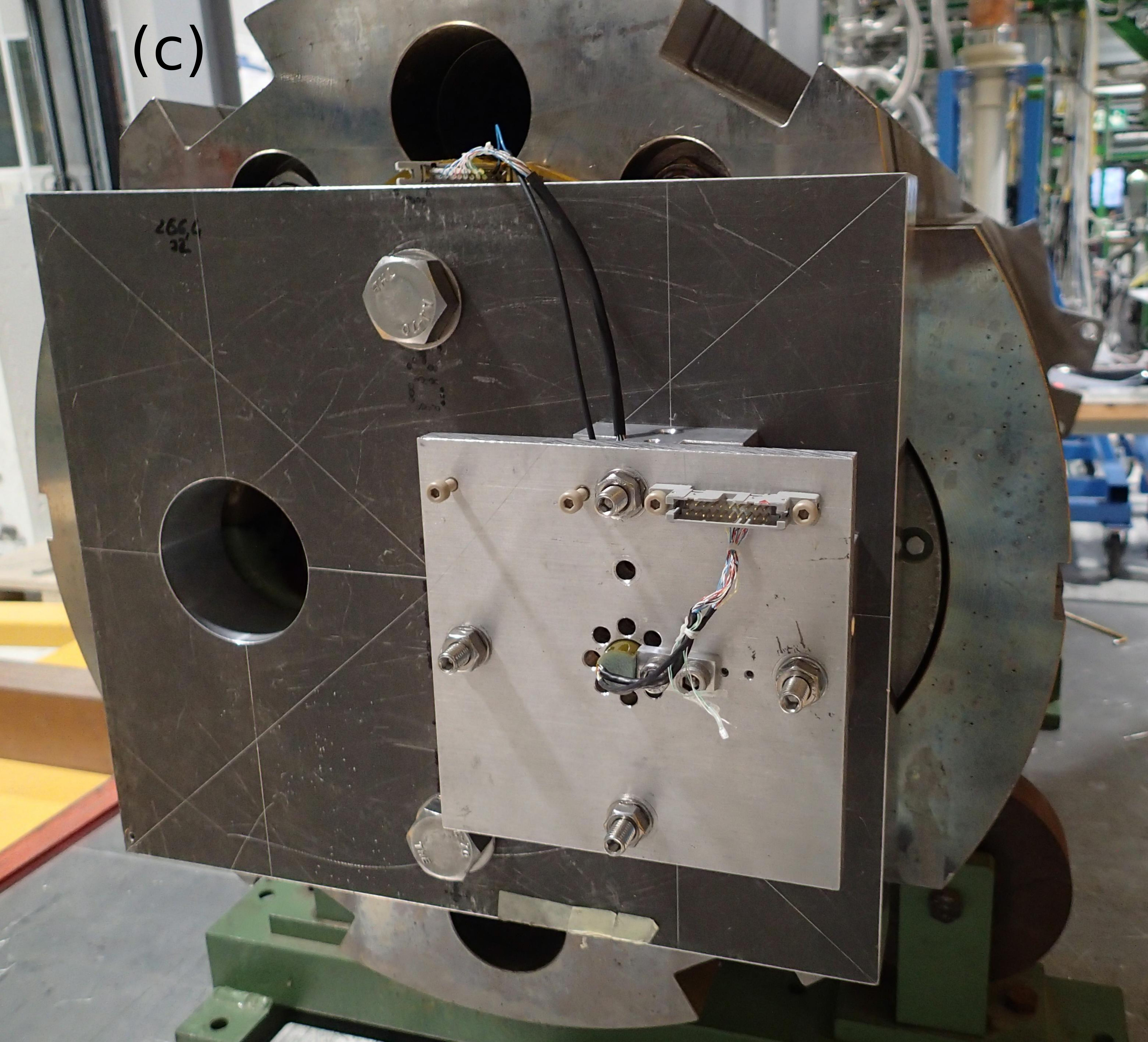}%
}
  \caption{The shield installed in one of the two apertures of the magnet. (a) Clamps to align the shield's position and orientation. (b) Delrin rod with the Hall sensors. (c) End plate to hold the shield in the magnet against the repulsive magnetic forces. }
  \label{fig:shield-in-magnet}
\end{figure*}

The  half  cylinders  were  clamped  between  a  bronze  tube  support
(ID/OD=18/41~mm) and  half-cylindrical aluminium  clamps, as  shown in
Fig.~\ref{fig:superconducting-shield-construction}.         Calibrated
high-sensitivity  Hall sensors  (Arepoc  HHP-NP)  were installed  into
slots of the aluminium clamps  to measure the external magnetic field,
with  a parallel  orientation.  The  same  type of  Hall sensors  were
mounted to  a delrin rod, which  was inserted into the  bronze support
tube.  These sensors were aligned both parallel and perpendicularly to
the external field.   The layout and the numbering scheme  of the Hall
sensors          is           shown          schematically          in
Fig.~\ref{fig:schematic-experimental-layout}.  The active  spot of the
external Hall sensors was about 2.6~mm  away from the outer surface of
the shield.   The sensitivity  of the  individual sensors  was between
150-200~mV/T at 4.2~K, allowing the measurement of fields below the mT
scale using  sensitive voltmeters.  The  sensors were driven  by 20~mA
(Current  Generator  type Keithley  6221)  connected  in series.   The
current and voltage measurement leads of the sensors were twisted wire
pairs, and the series connection was  done at an external patch panel,
thereby  avoiding that  the  series of  the sensors  acts  as a  large
inductive pickup  coil.  The voltage  of the  sensors was read  out by
digital multimeters  (Keithley 2000) and  recorded by a computer  at a
sampling rate  of 10~Hz.  The  wires of  sensor \#6 got  broken during
cool-down, and  this sensor  was therefore not  used in  the analysis.
Sensors \#6 and  \#7 are in the  full field region of  the magnet, and
measured equal values in earlier tests with an MgB$_2$ shield. Sensors
\#8 and \#9 are in the fringe field and measure lower values.

A thin pickup coil was installed  around the shield in the gap between
the      two      aluminium       clamps,      as      visible      in
Fig.~\ref{fig:superconducting-shield-construction}(a).  Unfortunately,
one half of the coil  was crashed between the half-cylindrical sheets,
and short-circuited to them.  This half  was replaced by the blue wire
taped  to the  outside surface  of the  clamps in  the midplane.   The
other, original  branch of the  coil is  hardly visible in  the figure
among the wires of the Hall sensors.   The purpose of this coil was to
pick up  sudden changes of  the external magnetic  field in case  of a
flux jump, measure its time difference  with respect to the signals of
the internal Hall sensors, and  ultimately to evaluate the feasibility
of this method as an early  diagnostics of flux jumps, to safely abort
the beam  before the  penetrating field  has fatal  consequences.  The
voltage measured at the two terminals  of the pickup coil was measured
by a  fast digital integrator (FDI  v3 \cite{Arpaia2006AFastDigital}),
and recorded by a computer at a sampling rate of 1~kHz.

The shield  was installed  into the  bore of a  spare LHC  MCBY dipole
corrector  magnet, as  shown  in Fig.~\ref{fig:shield-in-magnet}.  The
magnet  has  two  large   apertures  (70~mm),  powered  independently.
Without  the  shield this  magnet  creates  a high-quality  transverse
dipole  field, perpendicular  to its  axis. The  magnet has  a nominal
field of  2.5~T at  4.5~K and 72~A  \cite{LHCDesignReportVol1Ch8}, but
the achievable  field in the  presence of  the shield is  higher.  The
length of the shield was chosen  initially such that it extends beyond
the fringe fields of another,  short but large-aperture magnet at both
ends. Unfortunately,  this magnet  was not  available anymore  and the
magnetic length  of the MCBY  magnet (0.899~m) exceeded the  length of
the shield.   The shield  was therefore installed  asymmetrically into
the magnet, with one of its ends being outside of the fringe field, so
that eventual  effects due to the  shield's open end being  exposed to
the strong field  can be identified.  The full setup  was installed in
the Siegtal cryostat of the SM18  facility of CERN, and fully immersed
in liquid helium.

\section{Numerical methods}

Two different 2-dimensional finite element simulation models were used
to  reproduce   the  experimental   results.   (i)   Campbell's  model
\cite{Campbell2007ANewMethod}  is   a  static  model   which  directly
calculates  the approximative  steady state  of Bean's  critical state
model  \cite{Bean1964MagnetizationOfHighField}, obtained  by a  direct
ramp from a virgin state.  This  method is fast and therefore adequate
for  parameter  scans and  optimization.   It  is not  applicable  for
time-dependent phenomena,  such as relaxation, and  for magnetic field
ramps with different directions,  i.e. hysteresis simulations. For the
non-symmetrical   cases   extra  parameters   (the  values   of  the
vector-potential $A_z$ in the interior of the bulk superconductors) were
introduced  and  solved for  by  requiring  that the  current  density
integrated over  the cross  section of  each superconducting  piece be
zero,  as   described  in  \cite{Barna2017HighFieldSeptum}.    (ii)  A
time-dependent  eddy  current  simulation   using  the  power-law  E-J
characteristics   $E   =   E_0\cdot\left[J(B)/J_c(B)\right]^n$,   with
$E_0$=100~$\mu$V/m and $n$=100, typical  values used in the literature
\cite{Russenschuck2010FieldComputation}.  In  both cases  the critical
current  density of  the shield  material was  taken from  Figure~2 of
reference~\cite{Itoh1995CriticalCurrentDensity},  multiplied   by  the
NbTi filling factor of 0.36 (transport current parallel to the rolling
direction of  the sheet, 350~$^\circ$C~$\times$~672~h  heat treatment,
dashed line and open symbols).

\section{Results}

\begin{figure}
  \centerline{\includegraphics[width=\linewidth,height=\deffigheight]{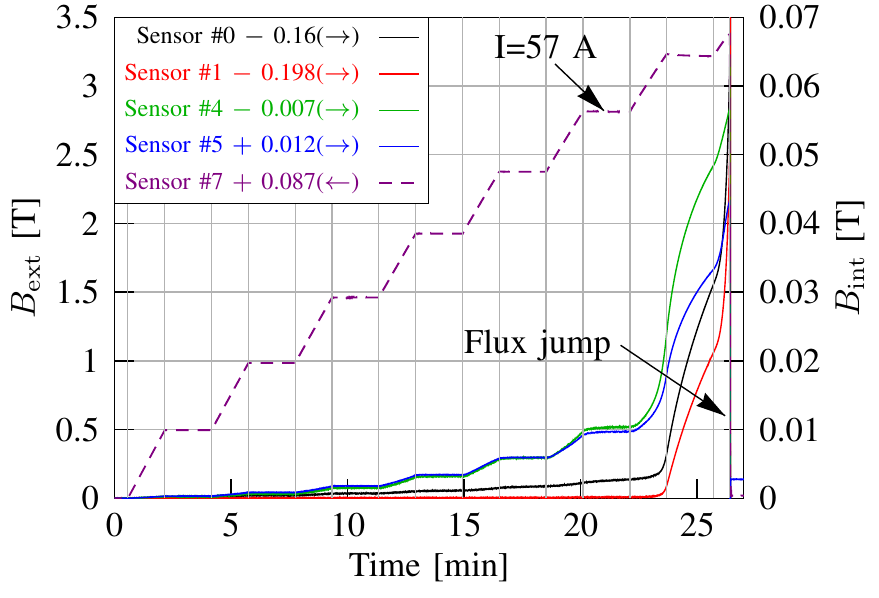}}
\caption{Color online. Magnetic field measured by the external (dashed lines) and
  internal (solid lines) Hall sensors, after offset (trapped field)
  subtraction (see text for details). The vertical lines indicate the
  start and end of the ramps. Plateaus correspond to integer multiples
of 9.5~A magnet current. Ramp rate is 0.1~A/s. Arrows in the legend indicate the vertical axis.}
\label{fig:cycle-1}
\end{figure}

After the installation and cool-down of the setup, initial tests of the
magnet, its  power supply  and quench  protection system  were carried
out,  which included  fast ramp-ups  of the  magnet current,  and fast
energy extraction. These have lead to flux jumps in the shield, or the
quench of the magnet, which in turn induced a flux jump in the shield.
The first measurement of the shield was carried out starting from this
state with trapped magnetic field.  Figure~\ref{fig:cycle-1} shows the
magnetic  field levels  after  the subtraction  of  the trapped  field
offset  (indicated  in the  legends).   A  strong penetration  of  the
changes  in the  external  field  started during  the  ramp after  the
6$^\mathrm{th}$ plateau  (starting at around 22~minutes).   It must be
noted that  this penetration is  smooth and relatively slow  yet.  The
avalanche-like sudden  collapse of the shielding  currents (flux jump)
occurred at  about 26  minutes.  This  triggered the  magnet protection
system  and  the magnet  current  was  very  quickly ramped  to  zero,
terminating the measurement cycle.

\begin{figure}
  \centerline{\includegraphics[width=\linewidth,height=\deffigheight]{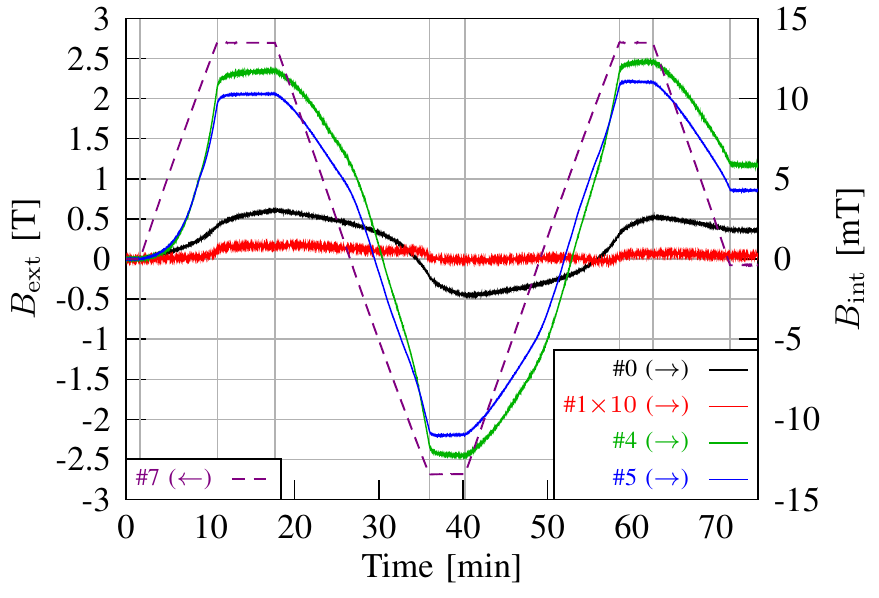}}
  \caption{Color online. Internal and  external magnetic  fields measured  during a
    cycle between  $\pm$55~A starting from  a virgin state.  The legends
    indicate scaling  factors. }
\label{fig:cycle-2-magnetic-cycle}
\end{figure}

In order to eliminate the trapped  field from the shield it was warmed
up above its critical temperature. The  signal of the Hall sensors was
monitored during the warm-up, and  clearly indicated the transition of
the shield to  normal-conducting state by the  sudden disappearance of
the trapped field.  The temperatures  shown by the sensors attached to
the  magnet were  around 50~K,  when cool-down  in zero  field started
again.  Even though electric heaters  were attached to the magnet, the
complete cycle  took almost 24~hours  due to  the long time  needed to
evaporate liquid  helium from the  large cryostat, and the  large heat
capacity of the 1.2~tons magnet.   Testing the ultimate shielded field
starting  from  a  virgin  state   would  have  led  to  another  full
penetration and  the loss  of another 24~hours  due to  the subsequent
thermal reset cycle.  Due to the limited time for the experiment, this
test was  omitted.  Subsequent measurements  from a virgin  state were
limited to a  magnet current of 55~A, slightly below  the value at the
last stable plateau (57~A)  in Fig.~\ref{fig:cycle-1}, hoping and then
finding that this is still below the penetration limit.  The following
results  indicate  therefore  only  a lower  limit  of  the  shielding
capability  of  the  shield.   Figure~\ref{fig:cycle-2-magnetic-cycle}
shows  the   magnetic  field  measurements  during   a  cycle  between
$\pm$55~A, starting from the virgin state.  At the highest current the
external  magnetic  field  measured   by  sensor~\#7  was  $\pm$2.7~T.
According to a 2D finite  element simulation of the experimental setup
(taking into  account the exact coil  and yoke geometry of  the magnet
and the shield) this corresponds to a magnetic field of about 3.1~T at
the shield's surface.  Among the internal Hall sensors (solid lines of
Fig.~\ref{fig:cycle-2-magnetic-cycle})  the  ones  with  perpendicular
orientation (\#4  and \#5) measured  by far  the largest field,  up to
12.5~mT.  This  corresponds to an attenuation  of $4.6\times 10^{-3}$.
Sensor \#0,  oriented parallel  to the  external field,  measured only
3~mT, corresponding  to an attenuation  of $10^{-3}$. This value  is 5
times more than that resulting from a 3D simulation assuming a perfect
diamagnet  shield.   Field leakage  of  parallel  orientation at  this
position is due to this sensor being  close (86~mm) to the open end of
the shield with a comparably large aperture (41~mm).  Inner sensor \#1
(at the  same axial position  as the  external sensor \#7)  measured a
magnetic field only below 0.1~mT,  which corresponds to an attenuation
of $4\times 10^{-5}$, already acceptable for the intended application.
The corresponding value in the  3D simulation with a perfect diamagnet
shield was zero within the precision of the simulation.  At the end of
the cycle both the external and the internal sensors show the presence
of trapped magnetic field.

\begin{figure}
  \centerline{\includegraphics[width=\linewidth]{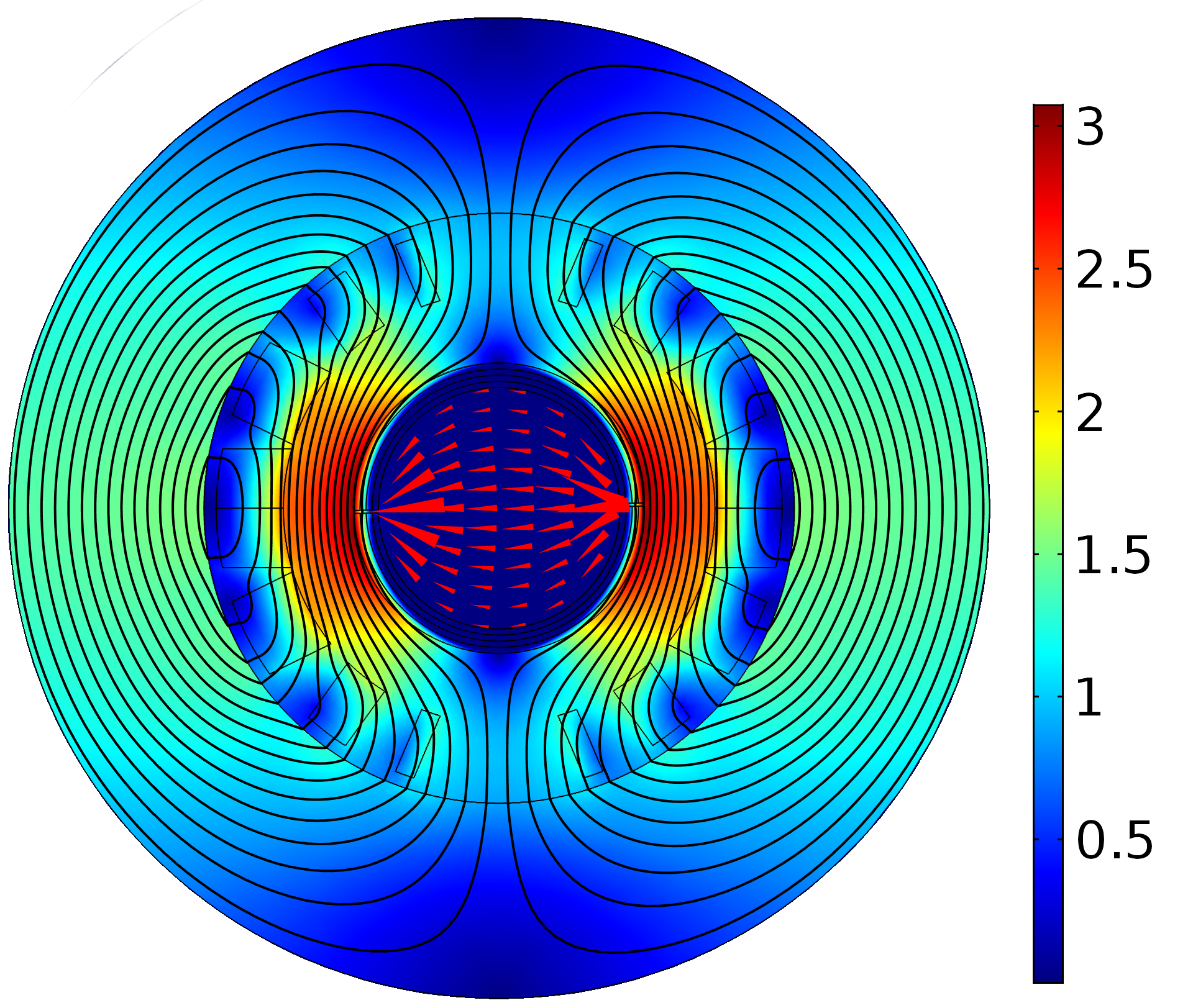}}
  \caption{Color online. Finite-element simulation of the cut shield geometry with
  improper alignment. Black lines and red arrows indicate the magnetic
  field.}
  \label{fig:cut-simulation}
\end{figure}

Figure~\ref{fig:cut-simulation} shows  the results of a  2D simulation
using Campbell's  method for the  experimental geometry with a  cut of
0.5~mm and rotation  of the shield by 1.5$^\circ$.  The dominant field
component is  perpendicular to the  external field: the  field levels
inside  the shield  are  15~mT  and 0.4~mT  in  the perpendicular  and
parallel directions, respectively.   This is a hint  that the observed
leakage magnetic field can be attributed to the actual shield geometry
with       unprecise       alignment,      as       suggested       in
Fig.~\ref{fig:cut-and-exp-layout}(b),  and  not  limited  by  material
performance.  An ideally arranged configuration [such as that shown in
  Fig.~\ref{fig:cut-and-exp-layout}(c)]  extending  safely beyond  the
fringe field of  the magnet would presumably perform at  least as well
as suggested by  sensor \#1, i.e.  with an attenuation  of better than
$4\times 10^{-5}$.

\begin{figure}
  \centerline{\includegraphics[width=\linewidth,height=0.9\deffigheight]{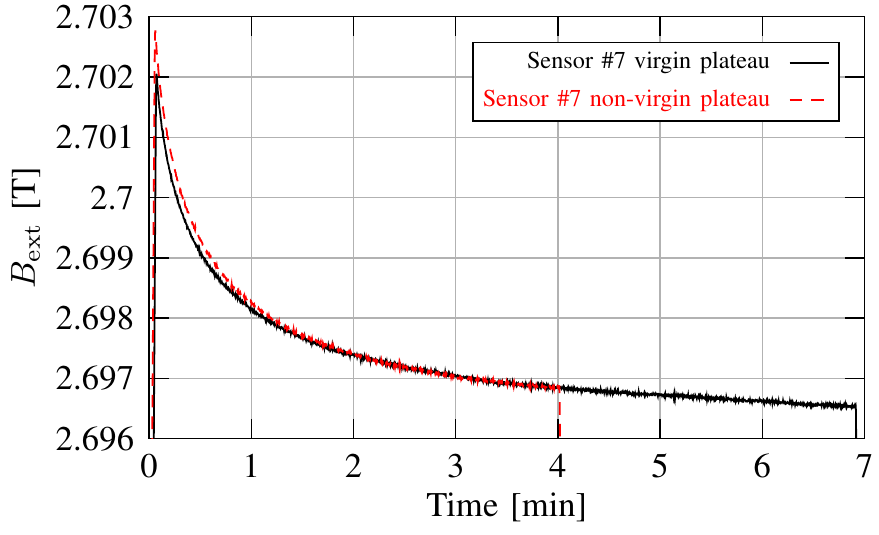}}
  \caption{ Comparison  of the field  measured by
    sensor \#7 on the two plateaus.}
\label{fig:cycle-2-relaxation-on-plateaus}
\end{figure}

Figure~\ref{fig:cycle-2-relaxation-on-plateaus} shows the magnetic field measured by the
external sensor \#7 on the two plateaus, shifted on the horizontal
axis to match their starting points. The behaviour is very similar in
the two cases. The relaxation is about 0.26\% over 7 minutes, and
saturates with time. 

\begin{figure}
  \centerline{\includegraphics[width=\linewidth,height=1.3\deffigheight]{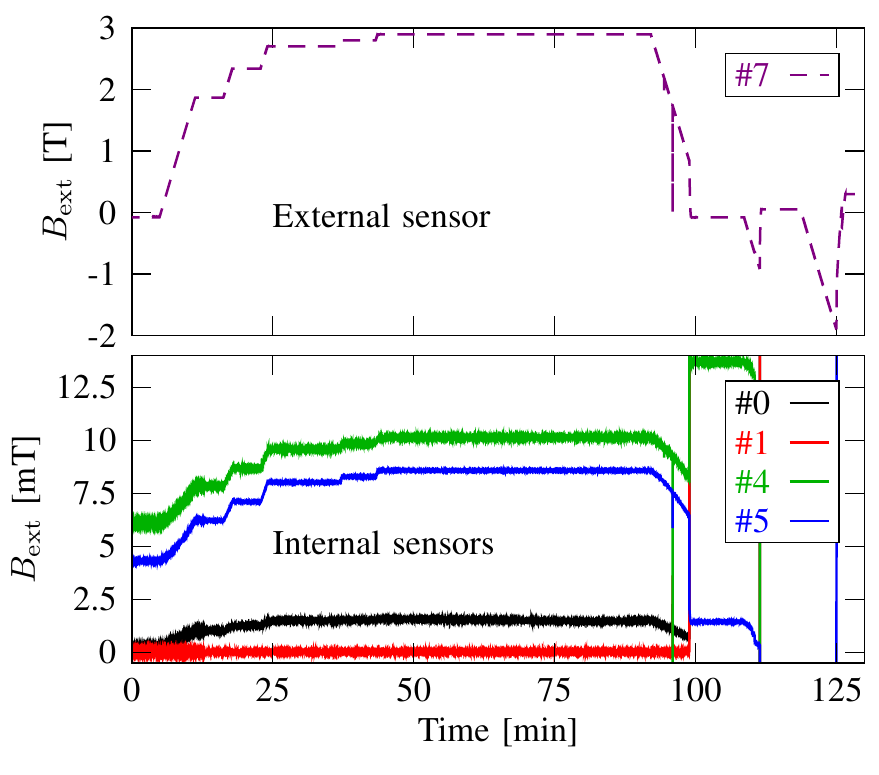}}
  \caption{Color online. Magnetic field measurements at 1.9~K.}
  \label{fig:1.9K}
\end{figure}

Following this  cycle, the shield was  cooled down to 1.9~K  without a
thermal reset  cycle. The  magnet current was  linearly ramped  to 38,
47.5, 55,  57, 59~A  in a  sequence.  Figure~\ref{fig:1.9K}  shows the
magnetic field measured inside and  outside the shield.  Small offsets
at the  beginning are due  to the trapped  field.  In contrast  to the
results at 4.2~K,  the inner sensors showed no creep  on the plateaus,
and  the maximum  variation of  their values  was significantly  less,
4.5~mT.   For sensor  \#0 this  variation was  1.5~mT, only  2.5 times
higher than the leakage field in  the 3D simulation assuming a perfect
diamagnet  shield.   These  reflect  the  smaller  relaxation  of  the
shielding currents  and the  higher value of  the critical  current at
lower  temperatures.   Rather  than  testing  the  ultimate  shielding
performance  of the  shield  at 1.9~K  which would  have  lead to  the
penetration of the  field into the shield's  interior (either smoothly
or as  a flux  jump), the magnet  current was ramped  down to  zero in
order to test the stability of the  shield during a full cycle. A flux
jump occurred  when reaching about  0.7~T outside the  shield. Further
attempts  to  ramp  up  the  magnet current  again  were  hindered  by
persisting   flux  jumps.    This  phenomenon   was  similar   to  the
observations with an MgB$_2$ shield  at 4.2~K in a similar measurement
\cite{BarnaFCCWeek2017}. In spite of the better shielding performance,
superfluid helium  temperatures are  therefore not applicable  for the
high-field septum concept.

\begin{figure*}
  \centerline{\includegraphics[width=0.95\linewidth,height=\deffigheight]{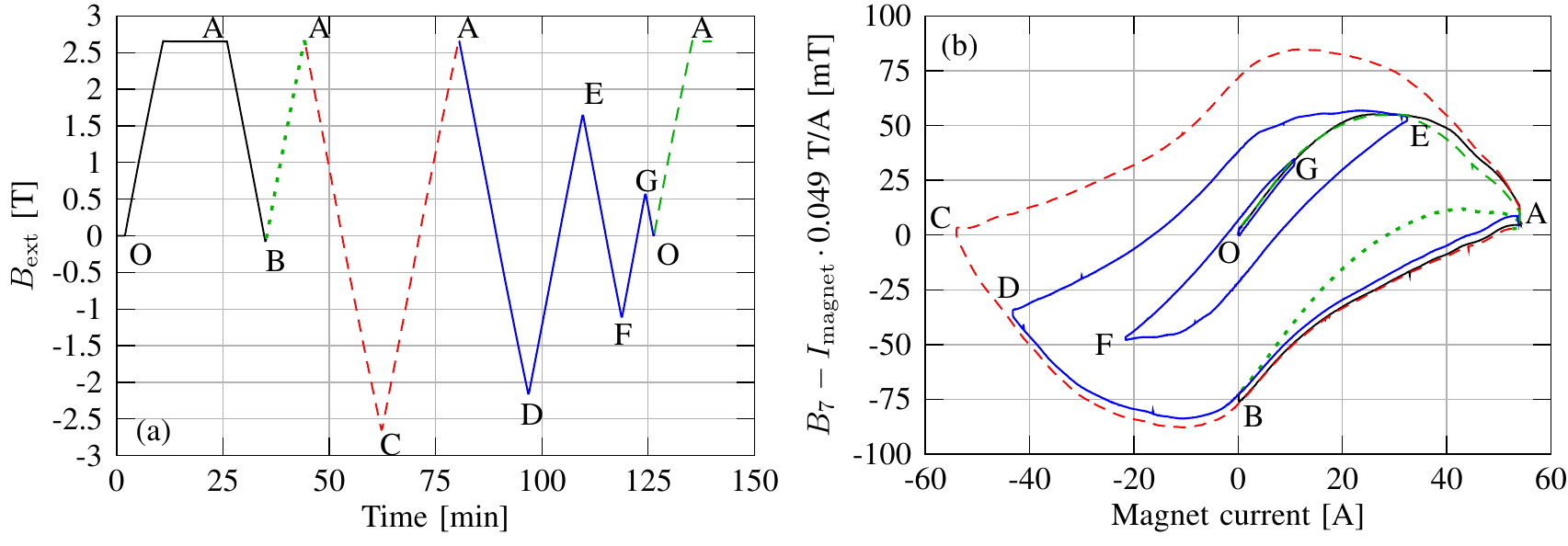}}
  \caption{Color online. Demagnetization of the shield. (a) External magnetic field
    measured by sensor \#7 as a function of time. (b) Deviation of the
    external magnetic field from a linear behaviour, as a function of
    magnet current.}
  \label{fig:demagnetization-exp}
\end{figure*}

\begin{figure*}
  \centerline{\hspace*{\fill}%
    \afig{width=0.4\linewidth,height=\deffigheight}{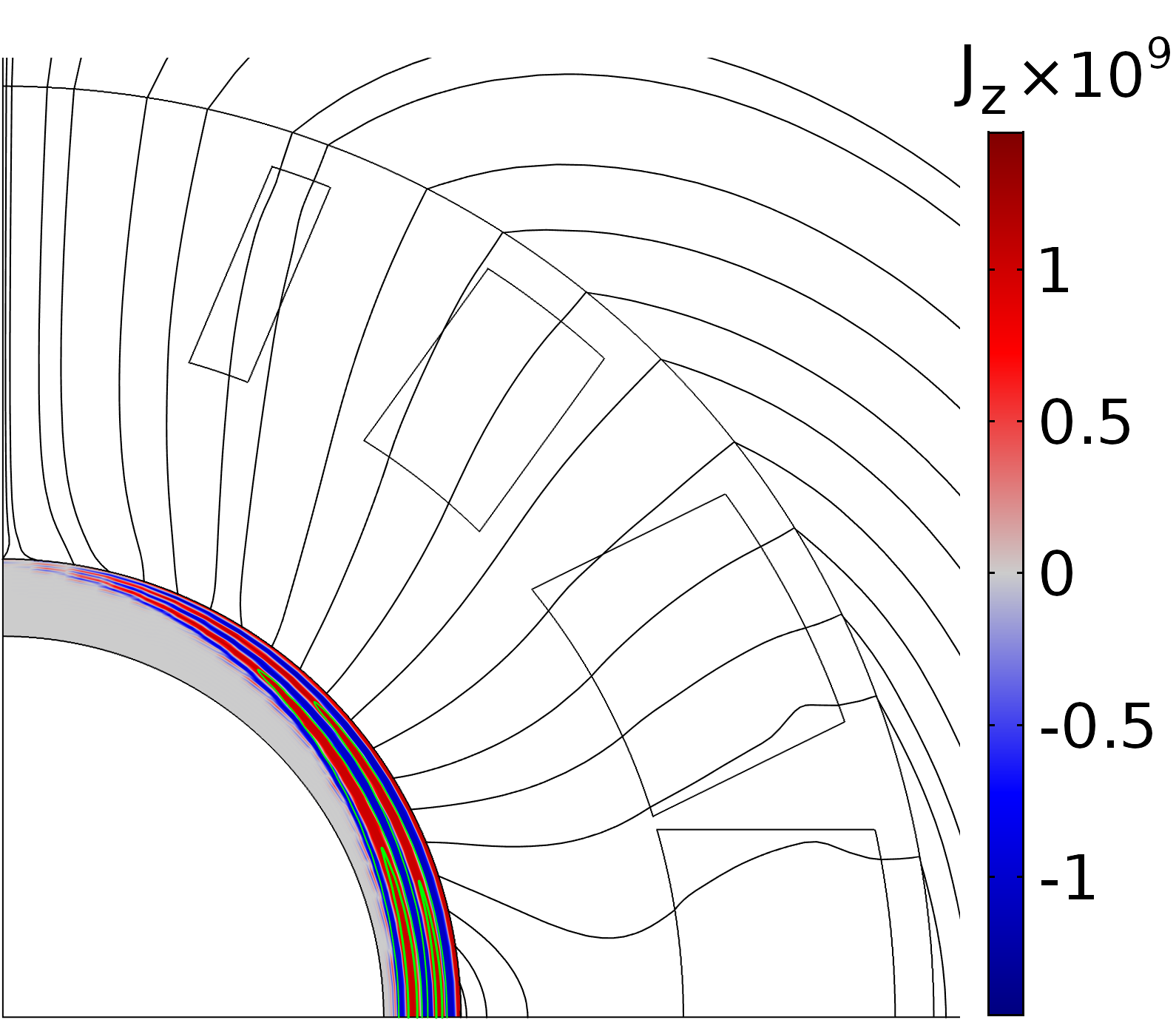}{7,7}{(a)}\hspace*{\fill}%
    \afig{width=0.5\linewidth,height=\deffigheight}{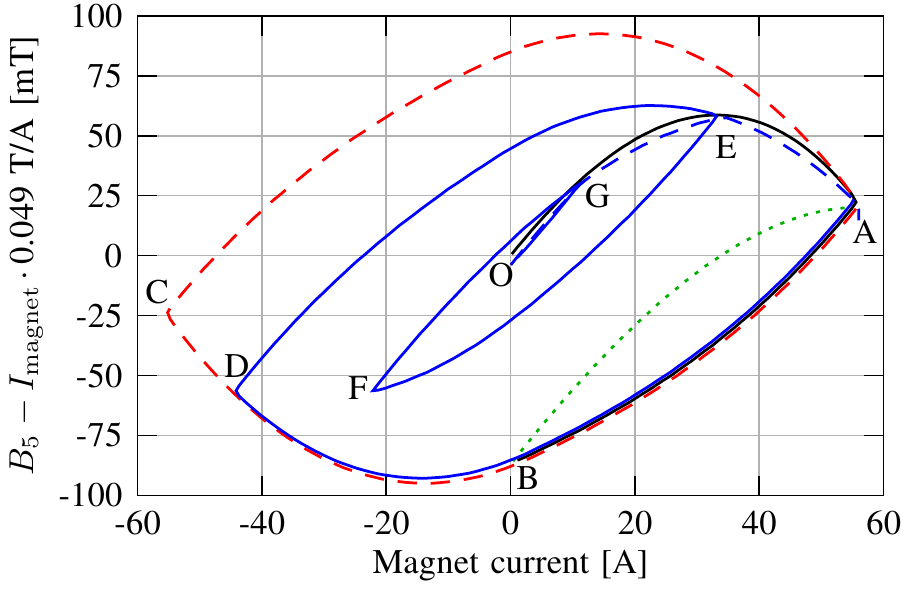}{20,13}{(b)}\hspace*{\fill}}
\caption{Color online. (a)  Simulated   pattern  of  the  magnetic   field  after  a
  degaussing  cycle. Color  scale  indicates  the persistent  currents
  $J_z$. Black and green lines  indicate the induction lines, color is
  only for better visibility. The  distribution of the induction lines
  was chosen  manually for  illustration purposes, their  density does
  not reflect actual field strength.  (b) Reproduction of the measured
  hysteresis   cycle   [Fig.~\ref{fig:demagnetization-exp}(b)]   by
  simulation}
\label{fig:demagnetization-sim}
\end{figure*}

A quantitative study of the effect  of trapped field for the realistic
configuration of  a SuShi septum  is clearly  beyond the scope  of the
present paper.   Here we only  demonstrate a possibility  to eliminate
this    effect    using    a     kind    of    `degaussing'    cycle
(Fig.~\ref{fig:demagnetization-exp}) at 4.2~K.   The  shield   started  from  a
virgin state after  a thermal reset. The magnet current  was ramped to
54~A and then back to zero.   This corresponds to the solid black line
$O$-$A$-$B$ in Fig.~\ref{fig:demagnetization-exp}(b).  At zero current
the trapped  field at the position  of sensor \#7 was  75~mT. A second
ramp to  54~A (green dotted line  $B$-$A$) had a trace  different from
the virgin  curve but reached the  final endpoint $A$ as  before. This
illustrates the  effect that exposures to  fields up to or  beyond the
highest  level  reached  before  erase the  magnetic  history  of  the
shield. A  double-ramp to  -54~A and 54~A  traced the  full, symmetric
hysteresis  loop (red  dashed line  $A$-$C$-$A$).  A  degaussing cycle
with alternating polarities and decreasing amplitudes (solid blue line
$A$-$D$-$E$-$F$-$G$-$O$) brought the shield back to the same effective
magnetic state $O$ as the starting  point.  The trace of the last ramp
to 54~A (green dashed line) seems  to slightly deviate from the virgin
curve,  but  reaches  the  same  endpoint  $A$.   The  phenomenon  was
simulated using the time-dependent method  and the same magnet current
profile as that used in the  experiment, except on a shorter timescale
(10.8~A/s ramp rate, no plateaus),  to make the simulation run faster.
Figure~\ref{fig:demagnetization-sim}(a)  shows  the  fine  alternating
pattern  of  the persistent  currents  and  magnetic field  after  the
degaussing  cycle.  The  majority of  the induction  lines are  closed
within the shield.  The stray field at the position of the Hall sensor
is negligible. These results illustrate that the magnetic state of the
shield is  reset only at  the effective level which,  nevertheless, is
sufficient  for  the  intended  application.   The  microscopic  field
pattern still carries information about the shield's magnetic history.
Figure~\ref{fig:demagnetization-sim}(b)  shows   the  same  hysteresis
loops  as  in Fig.~\ref{fig:demagnetization-exp}(b).  
The last  ramp to  54~A (blue  dashed line)  deviates from  the virgin
curve, but touches the previous  endpoints of the degaussing cycle $G$
and $E$.  This  is due to the fact that  the persistent current layers
are erased in  a sequence when the field penetrates  again through the
wall of the shield. A magnetic  state identical to a previous one will
be reached  when a complete  persistent current layer is  erased.  The
differences  between  the  simulation  and  experimental  results  are
probably due to the faster ramp rates in the simulation (less time for
relaxation, higher instantaneous  induced currents), the approximative
nature and the non-optimized parameters of  the E-J power law, and the
difference between  the $J_c(B)$ curve  used in the simulation  and in
reality.  The simulation  nicely describes  the experimental  findings
qualitatively    and    helps    to    understand    the    underlying
phenomena. Since a time-dependent nonlinear eddy current simulation is
computationally  expensive, optimization  of  the  parameters was  not
attempted.

\begin{figure*}
  \centerline{\includegraphics[width=\linewidth,height=\deffigheight]{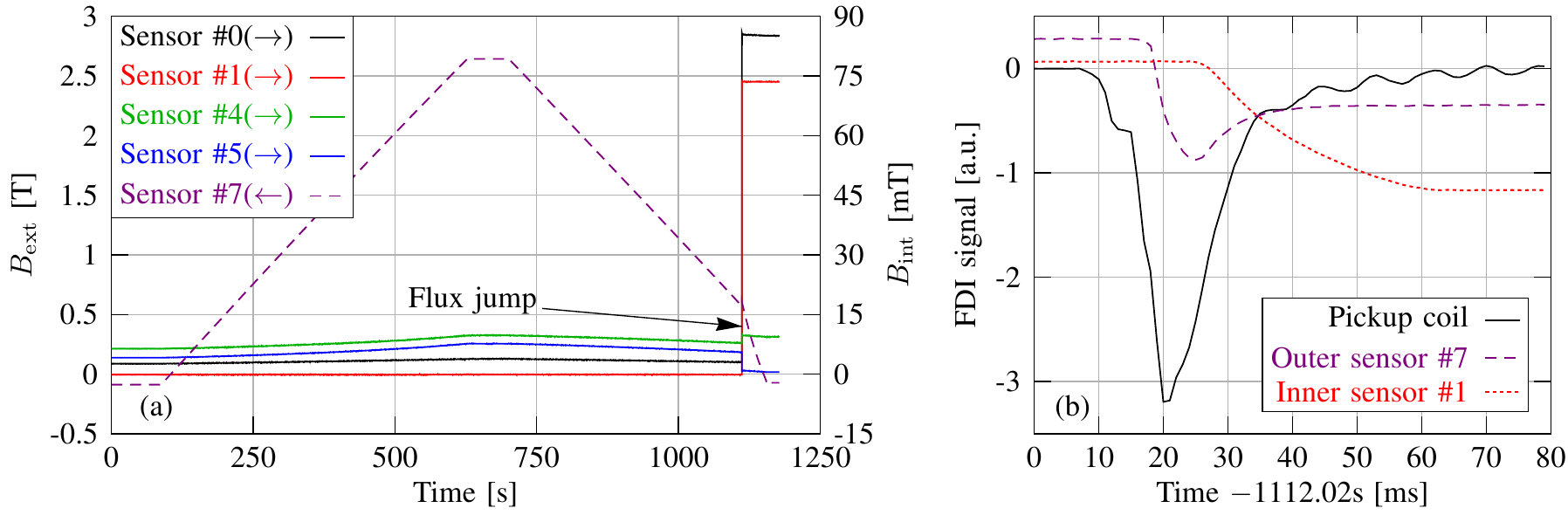}}
  \caption{Color online. Timing measurements of a flux jump at 1.9~K}
  \label{fig:cycle-6-flux-jump}
\end{figure*}

The slow data acquisition rate (10~Hz) of the Hall sensors read out by
the multimeters  did not  allow to  measure eventual  time differences
below 100~ms  between the signals of  the Hall sensors and  the pickup
coil in case of a flux jump. For this measurement the Hall sensors \#1
(inside the shield) and \#7 (outside the shield) were connected to two
other channels of  the FDI with a  gain of 2 to obtain  the same 1~kHz
sampling rate as for the pickup coil.  At 4.2~K the magnet current was
ramped to  57~A with  a ramp rate  of 0.5 and  1~A/s.  The  shield was
stable and  no flux jump occurred.   When higher ramp rates  were set,
the power  converter tripped.  A  flux jump provoked by  higher magnet
currents  would have  been preceded  by  a smooth  penetration, as  in
Fig.~\ref{fig:cycle-1} and in \cite{Itoh1993MagneticShielding}.  Given
the  apparent  stability of  the  shield  against  flux jumps  at  the
intended field levels,  in a realistic scenario an  eventual flux jump
would be  caused by an  external perturbation, like  energy deposition
due to beam  loss. In this case  a flux jump would  occur suddenly and
directly  from a  perfectly shielding  state.  In  order to  trigger a
similar   situation,   the   shield   was  cooled   down   to   1.9~K.
Figure~\ref{fig:cycle-6-flux-jump}(a)  shows  the  usual  Hall  sensor
curves as  a function of  time. As expected,  a flux jump  occurred at
$t$=1112~s.   The   signals  recorded  by   the  FDIs  are   shown  in
Figure~\ref{fig:cycle-6-flux-jump}(b).  A clear peak  in the signal of
the pickup coil precedes the peak measured by the external Hall sensor
by about 10~ms, and the departure of the internal Hall sensor's signal
by about  15~ms.  This time  interval seems to  be safe to  trigger an
emergency beam abort in the ring.  Since the pickup coil encircles the
whole shield, its inductive signal  records flux jumps starting at any
point along the shield. In fact,  the recorded shape of the peak might
indicate flux jumps starting at  two different locations, with a small
time  difference, although  this statement  is rather  speculative. In
contrast, the Hall sensors are recording magnetic field levels at well
defined spots  inside and on  the outer  surface of the  shield.  Time
differences  between  the pick-up  coil  and  the Hall  sensors  might
therefore be a purely geometrical effect, caused by the propagation of
the instability along the axis of  the shield.  However, since the two
recorded  Hall sensors  are at  the same  longitudinal positions,  the
observed time  difference between their  signals can be  attributed to
the retarding  effect of the eddy  currents induced in the  shield and
its bronze and aluminium  support structure.  The quantitative results
are  therefore clearly  a function  of the  specific geometry  and the
amount of conductor material around the circulating beam, which should
be maximized in the final design.

\section{Conclusions and outlook}

The Future Circular Collider project  is seeking for novel concepts to
manipulate  its  proton beam  of  unprecedented  energy.  One  of  the
challenges is  the construction of  a high-field septum magnet  with a
field of at  least 3~T, and an apparent septum  thickness below 25~mm.
One of the proposals is to  realize this device using a combination of
a passive superconducting shield and a special superconducting magnet,
nicknamed  as a  SuShi septum.   This paper  reported on  the magnetic
shielding properties of a candidate superconducting shield material, a
0.8~mm  thick  NbTi/Nb/Cu  multilayer sheet.   A  cylindrical  shield,
constructed  from 4+4  layers of  this material,  in the  form of  two
half-cylinders, with  a total wall  thickness of 3.2~mm  could support
3.1~T on  its surface,  in accordance  with the  results of
  \cite{Itoh1993MagneticShielding},    even   though    the   shield's
  construction was  different.  We estimate  that with 5
layers and  a total thickness of  4~mm only, the shield  could support
3.2~T  with  a safety  margin,  the  current  value  used in  the  FCC
conceptual design  report for this  type of device.  Together  with an
additional  support of  11~mm thickness  and the  beam pipes  and beam
screens the apparent septum thickness would be below the target value.
The shield  material is ductile and  easy to form and  handle.  It was
stable against spontaneous flux jumps  at 4.2~K, and survived magnetic
cycles between opposite polarities  without flux jumps. Relaxations of
the  shielding  currents  are  at  a  tolerable  level,  and  we  have
demonstrated a ``degaussing''  method to eliminate the  effects of the
field trapped in  the shield's thick wall after  high field exposures.
Even though the  shield's performance was better at 1.9~K  in terms of
shielding efficiency  and relaxation rates, frequently  occurring flux
jumps make  this temperature inapplicable. The  observed properties of
the material make it an ideal candidate for the realization of a SuShi
septum magnet. Unfortunately the material is a discontinued product of
Nippon Steel Ltd., and its availability is not clear even on the short
term. The material  for the reported tests was purchased  from a small
remaining   stock   of   semi-finished  products   of   the   company,
post-processed to the final thickness  and specifications by a private
company  in  Japan.   If  the  material  can  be  produced  in  larger
quantities, the unit cost is expected to be reduced.

MgB$_2$, another  candidate material,  also demonstrated  an excellent
shielding  performance in  a similar  test.  It  supported 3~T  on its
surface  with a  wall  thickness of  8.5~mm,  perfectly shielding  its
interior \cite{BarnaFCCWeek2017}. However, it suffered from flux jumps
when the  external field  was ramped  down to  zero. This  material is
relatively cheap and easy to produce, and if the latter problem can be
solved,  it  provides  an  alternative to  the  NbTi/Nb/Cu  multilayer
sheet.  Flux jumps  are  one  of the  most  important  issues of  this
concept.  Stability   against  this  phenomenon  requires     careful
manufacturing and processing of the  material, and each shield must be
tested to be ``flux jump safe'' before assembly into the setup.

Encouraged by these positive test results,  a study is now underway to
design and  optimize a fully  fledged demonstrator prototype,  using a
canted  cosine  theta-like  magnet   and  a  half-moon  shaped  shield
\cite{BarnaFCCWeek2018}.  Besides the  demonstration of the achievable
maximum field  strength in  a realistic configuration,  this prototype
would  create a  homogeneous field  outside the  shield, and  it would
allow the measurement of the field quality in the high-field region.

\section*{Acknowledgements}
The authors would  like to express their gratitude to  the CERN TE-ABT
group, the  CERN  SM18  team, the  CERN  magnetic  measurements  group
TE-MSC-MM, Akira Yamamoto, Ikuo  Itoh, M\'arta Bajk\'o, Ranko Ostojic,
Fr\'ed\'eric  Rougemont, Yannick  Thuau.   This  project has  received
funding from  the FCC Study  Group, the European Union's  Horizon 2020
research  and innovation  programme  under grant  agreement No  730871
(ARIES),  and from  the Hungarian  National Research,  Development and
Innovation Office under grant \#K124945.

\bibliographystyle{IEEEtran}
\bibliography{extrareferences,references}

\end{document}